\documentclass{pos}

\usepackage{bm}

\title{Timelike formfactors of pion, kaon, and proton at large momentum transfers }

\ShortTitle{Timelike formfactors of pion, kaon, and proton at large momentum transfers }

\author{\speaker{Kamal Seth}\\
        Northwestern University\\
        E-mail: \email{kseth@northwestern.edu}}

\abstract{Form factors of the proton, pion, and kaon for large timelike momentum transfers have recently been measured with precision.  The results and future prospects are discussed.}

\FullConference{8th Conference Quark Confinement and the Hadron Spectrum \\
		 September 1-6 2008\\
		 Mainz, Germany}

\begin{document}

\section{Introduction}

Electromagnetic form factors of a hadron are the most direct link to the structure of the hadron in terms of its constituents.  They describe the coupling of a photon with a certain four--momentum to the distribution of charges and currents in the hadron.

The four--momentum transfer $Q^2$ in the collision of two particles with four-momenta $p_1$ and $p_2$ can be positive or space-like (in scattering) or negative or time-like (in annihilation/production).

\vspace*{-15pt}
\begin{center}
\begin{tabular}{cc}
\includegraphics[width=2.3in]{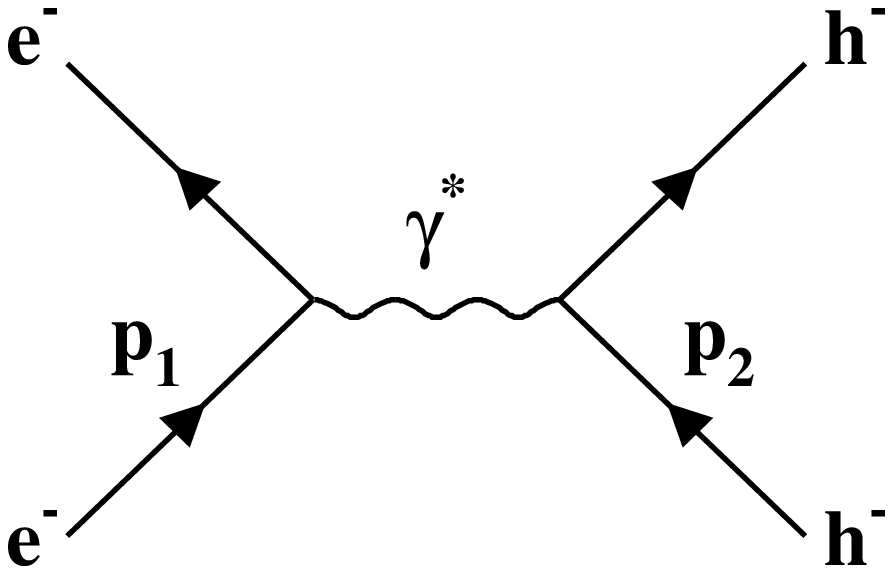}
\vspace*{-18pt}
&
\includegraphics[width=2.3in]{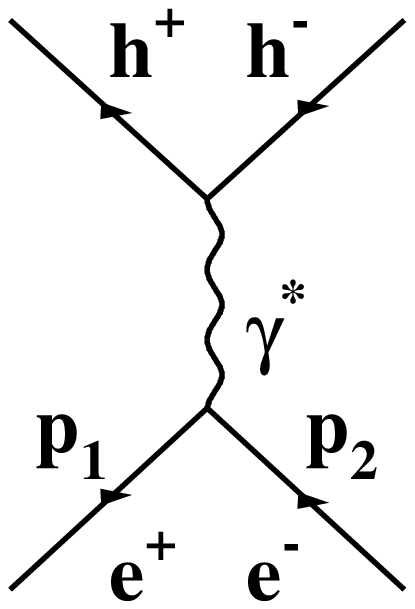}
\\
Scattering, Spacelike
&
Annihilation, Production
\\
positive $Q^2=t$
&
negative $Q^2=s$
\end{tabular}
\end{center}

 The form factor measurements done at SLAC and JLab with electron beams scattered from targets of $p$, $d$, ..., etc., and for electroproduction of pions (essentially electron scattering from the pion cloud) are exclusively for spacelike momentum transfers.  They require fixed targets, and are extremely difficult, if not impossible, to do for measuring space-like form factors of mesons at large momentum transfers; meson targets just do not exist!

Timelike form factor measurements for any hadron can be done with $e^+e^-$ annihilation, and for the special case of protons by $p\bar{p}$ annihilation.

Note: form factors are analytic functions of $Q^2$.  The Cauchy theorem alone guarantees that
$$F(Q^2,\textrm{timelike}) \stackrel{\atop Q^2\to\infty}{\longrightarrow} F(Q^2,\textrm{spacelike})$$

\section{ Cross Sections for Time-like Momentum Transfers}

For protons, there are two form factors, Pauli and Dirac Form Factors, or more familiarly, the magnetic $G_M(s)$ and the electric $G_E(s)$ form factors, and the cross section $e^+e^-\to p\bar{p}$~~is
$$\sigma_0(s) = \frac{4\pi\alpha^2}{3s}\beta_p \left[ |G_M^p(s)|^2 + \frac{\tau}{2}|G^p_E(s)|^2\right]$$

At large momentum transfers separation between $G_M(s)$ and $G_E(s)$ is very difficult, and the results which are generally reported assume $G_E(s)=0$, or $G_E(s)=G_M(s)$. 

For pions and kaons, both of which have spin 0, there is no magnetic contribution, and only the electric form factor $F(s)$ exists.  In this case the cross section for $e^+e^-\to m^+m^-$ is
$$\sigma_0(s) = \frac{\pi\alpha^2}{3s}\beta_m^3|F_m(s)|^2$$

The quark counting rules of pQCD predict that the baryon form factors are proportional to $Q^{-4}$ (or $s^{-2}$) and the meson form factors are proportional to $Q^{-2}$ (or $s^{-1}$), so that $ (d\sigma/d\Omega)_{\mathrm{proton}} \propto s^{-5}$, and $(d\sigma/d\Omega)_{\mathrm{meson}} \propto s^{-3}$, i.e., the cross sections fall very rapidly with increasing c.m. energy, and it becomes very difficult to measure any form factors at large momentum transfers. For example, $\sigma(e^+e^-\to p\bar{p})\approx1~\mathrm{pb}$ at $s=Q^2=13.5~\mathrm{GeV}^2$.  At $s=20~\mathrm{GeV}^2$ one expects to drop down by a factor $\sim7$, to $\sim150~\mathrm{fb}$.

Prior to the Fermilab (E760/E835) measurements in 1993/2003 of the timelike form factors of the proton by the reaction $p\bar{p}\to e^+e^-$, the data were sparse, had large errors, and were confined to $|Q^2|<5,7~\mathrm{GeV}^2$.  The Fermilab measurements obtained $G_M(|Q^2|)$ for four $|Q^2|$ between 8.9 and 13.11 $\mathrm{GeV}^2$~\cite{andreotti}.  As the solid curve in Fig.~1 shows, while $Q^4G_M(|Q^2|)$ was found to vary as $\alpha^2(\mathrm{strong})$, the value of the timelike form factor was \textbf{found to be twice as large} as the spacelike form factor, i.e., $R\equiv G_M(\mathrm{timelike})/G_M(\mathrm{spacelike})\approx2$.

Many theoretical attempts to explain $R\approx2$ using conventional models of the proton (the Mercedes star model) were made.  All were unsucessful.  This led Kroll and collaborators to propose the diquark--quark model of the nucleon.  While this model has at least two extra parameters, it did succeed in explaining both spacelike and timelike $G_M$, and $R\approx2$ quite nicely.

On the experimental side, there were new measurements of $G_M(p)$ using the $e^+e^-\to p\bar{p}$.  At Cornell we made a measurement of $G_M(p)$ at $|Q^2|=13.5~\mathrm{GeV}^2$~\cite{pedlar}, BES made direct measurements at ten values of $|Q^2|=4-9.4~\mathrm{GeV}^2$~\cite{ablikim}, and BaBar made measurements using ISR from $\Upsilon(4S)$ for $|Q^2|=3.6-20.3~\mathrm{GeV}^2$, albeit with large errors~\cite{aubert}.  All these measurements gave consistent results and confirmed $R\approx2$.  BaBar went a step farther, and derived $G_E/G_M$, though with even larger errors.

\begin{figure}[!tb]
\begin{center}
\includegraphics[width=2.9in]{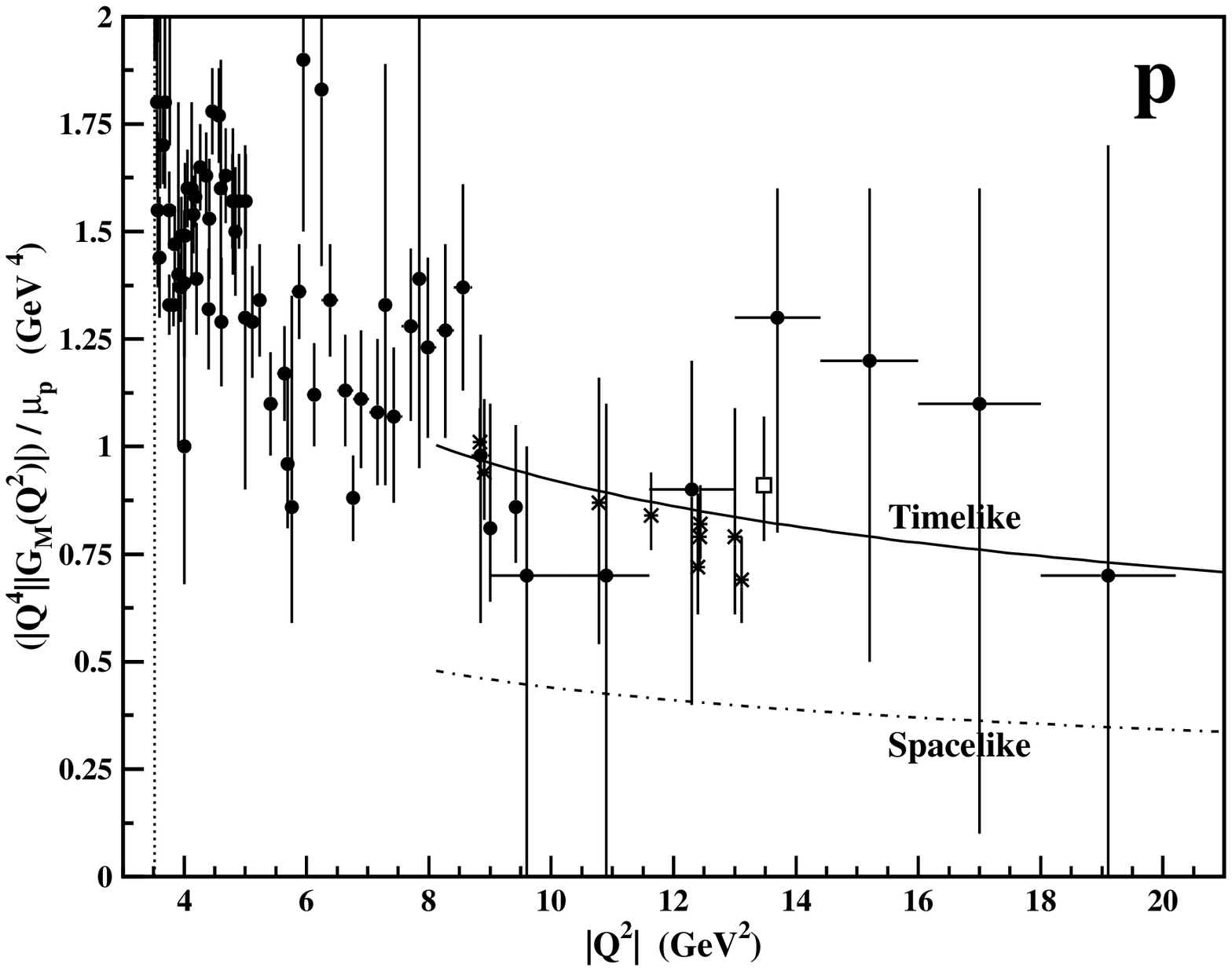}
\includegraphics[width=2.9in]{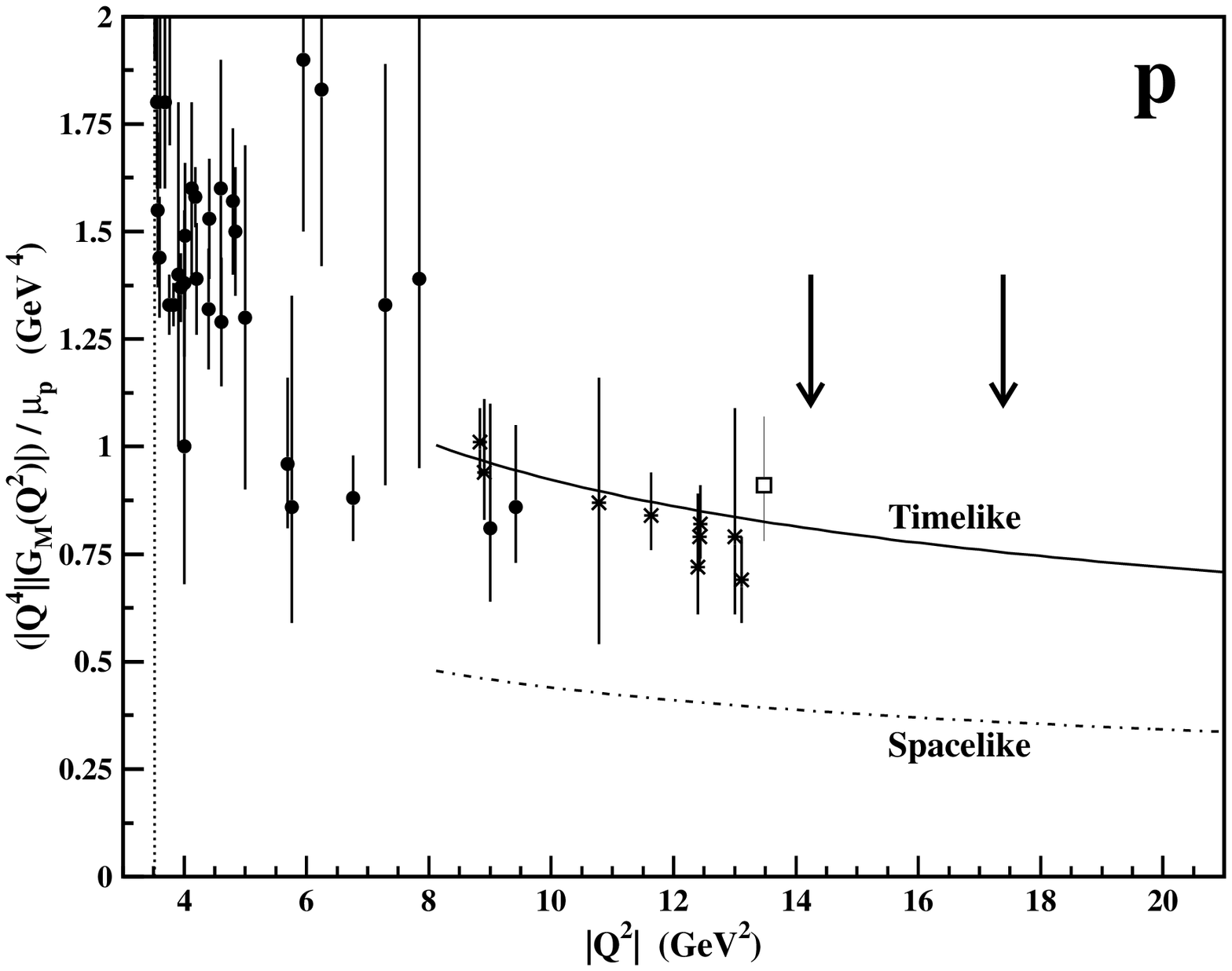}
\end{center}

\caption{(Left) World data on timelike form factors of the proton.  The points with large errors are ISR results from BaBar. (Right) Same as left, with BaBar results removed for clarity.  Arrows mark $|Q^2|=14.2$ and 17.4~GeV$^2$ at which new results are expected from CLEO.}
\end{figure}

\section{Form Factors of Pions and Kaons}

Mesons represent much simpler systems than baryons; two quark systems are expected to be easier to understand than three quark systems.  Indeed the now-classic debate about when $|Q^2|$ is large enough for the validity of pQCD took place in the 1980s between Brodsky and collaborators on one side and Isgur and Llwellyn Smith on the other side.  It was based on extremely limited and poor quality data for pion form factors, especially in the large $|Q^2|$ region which was the subject of the entire debate.  Recently, this experimental situation has changed drastically, mainly because of the measurements made by CLEO.

\begin{figure}[!tb]
\begin{center}
\includegraphics[width=2.9in]{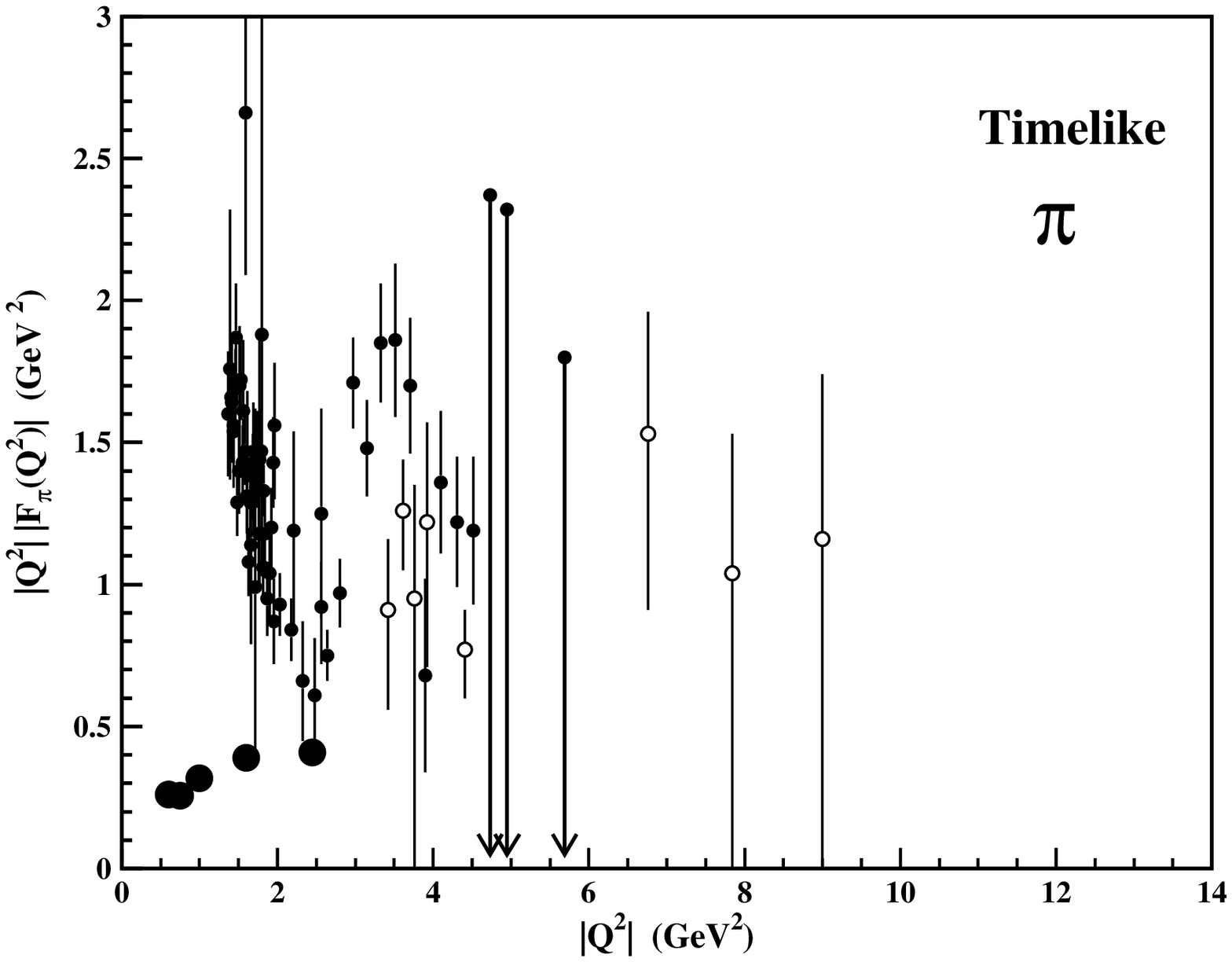}
\includegraphics[width=2.9in]{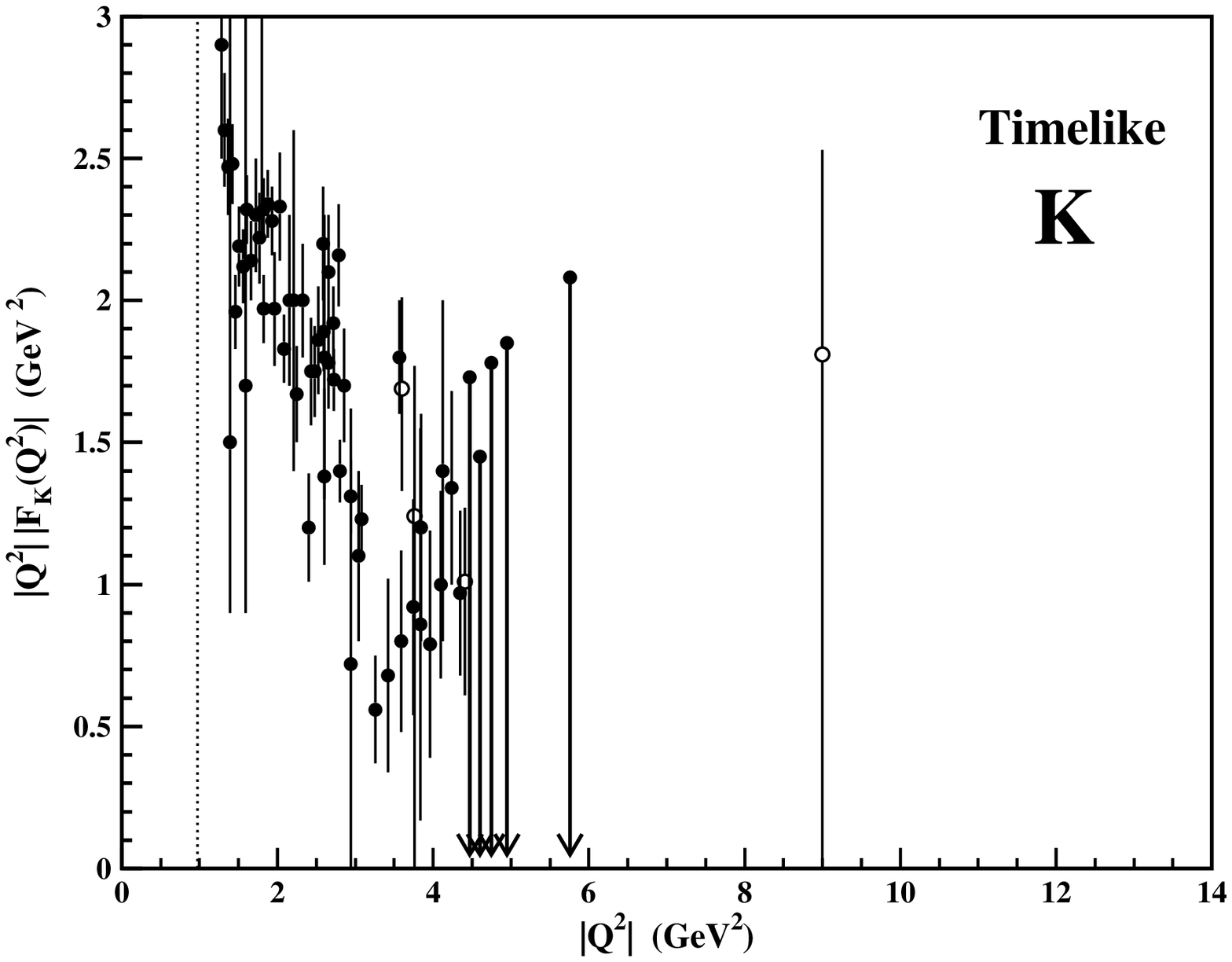}

\end{center}
\caption{Pion and kaon timelike form factors prior to the CLEO measurements.}
\end{figure}

\section{CLEO Measurements of Pion and Kaon Form Factors}

The CLEO measurements were made with the CLEO-c detector using 20.7 pb$^{-1}$ of $e^+e^-$ data taken at $\sqrt{s}=3.671~\mathrm{GeV}$, i.e., 15 MeV below the $\psi'$ resonance.  The data were originally taken for background studies for the $\psi'$ decays which were being studied.  It is ironic that these background studies have provided the world's best measurements of pion and kaon form factors.

To illustrate the formidable problem of backgrounds, let me jump a bit ahead to tell you that the CLEO measured form factor cross-sections at 3.67 GeV turn out to be $\sigma(e^+e^-\to \pi^+\pi^-)\approx8~\mathrm{pb}$, and $\sigma(e^+e^-\to K^+K^-)\approx4~\mathrm{pb}$.  The corresponding background cross-sections are: $\sigma(e^+e^-\to e^+e^-) \approx 130~\mathrm{nb}$, $\sigma(e^+e^-\to \mu^+\mu^-) \approx 5~\mathrm{nb}$, $\sigma(e^+e^-\to h\overline{h}) \approx 10~\mathrm{nb}$, i.e., $10^3$ to $10^5$ times larger than the form factor cross-sections to be measured.  

To reject backgrounds at this level one has to use everything at one's disposal.  This is what was done to identify $p\bar{p}$, $\pi^+\pi^-$, and $K^+K^-$.  Total observed pair energy, energy loss in the calorimeter, identification by the RICH detector, all were used to identify $14\pm5$~$p\bar{p}$, $26\pm5$~$\pi^+\pi^-$ and $82\pm10$~$K^+K^-$~events to obtain~\cite{pedlar}\vspace{8pt}\\
\hspace*{1.4in}PROTON: $|Q^4|G^p_M(|Q^2|=13.48~\mathrm{GeV}^2)= 0.91\pm0.16\pm0.04~\mathrm{GeV}^2$\\
\hspace*{1.4in}PION: $|Q^2|F_\pi(|Q^2|=13.48~\mathrm{GeV}^2)=1.01\pm0.11\pm0.07~\mathrm{GeV}^2$\\
\hspace*{1.4in}KAON: $|Q^2|F_K(|Q^2|=13.48~\mathrm{GeV}^2)=0.85\pm0.05\pm0.02~\mathrm{GeV}^2$
$$F_\pi(13.48~\mathrm{GeV}^2)/F_K(13.48~\mathrm{GeV}^2)=\bm{1.19\pm0.07}$$

The pion and kaon form factors were the \textbf{world's first} measurements of the form factors of any mesons at this large a momentum transfer, and with precision of this level, $\pm13\%$ for pions and $\pm6\%$ for kaons~\cite{pedlar}.  The results are shown in the figure along with the old world data, and arbitrarily normalized curves showing the pQCD predicted variation of $|Q^2|F_\pi$ and $|Q^2|F_K$ with $\alpha_S$.

In the figures, form factors at $|Q^2|=|M(J/\psi)|^2$ are also shown.  These are not from direct measurements, but are based on the argument of Milana~et~al.~\cite{milana}.  that
$$\frac{\mathcal{B}(J/\psi\to\pi^+\pi^-)}{\mathcal{B}(J/\psi\to e^+e^-)}=2F_\pi^2(M^2_{J/\psi})\times\left(\frac{p_\pi}{M_{J/\psi}}\right)^3$$
They thus obtained $|Q^2|F_\pi(|9.6~\mathrm{GeV}^2|)=0.94\pm0.06~\mathrm{GeV}^2$

The argument was extended by us to $J/\psi\to K^+K^-$ decay \cite{seth} to obtain
$$|Q^2|F_K(9.6~\mathrm{GeV}^2)=0.81\pm0.06~\mathrm{GeV}^2,$$
Both $F_\pi(9.6~\mathrm{GeV}^2)$ and $F_K(9.6~\mathrm{GeV}^2)$ so obtained are in remarkably good agreement with our measured values at 13.48~GeV$^2$.  We also note that 
$$F_\pi(M^2_{J/\psi})/F_K(M^2_{J/\psi})=\bm{1.16\pm0.27},$$
so obtained is also in excellent agreement with the above result of the CLEO measurement.

\section{Future Prospects}

As mentioned earlier, it is a regrettable face that none of the timelike form factors described here were obtained from dedicated measurements.  They result from exploiting background and off--resonance measurments.  So let us see if we can exploit other non--dedicated measurements, for example measurements at unbound charmonium resonances.

\begin{figure}[!tb]
\begin{center}
\includegraphics[width=2.9in]{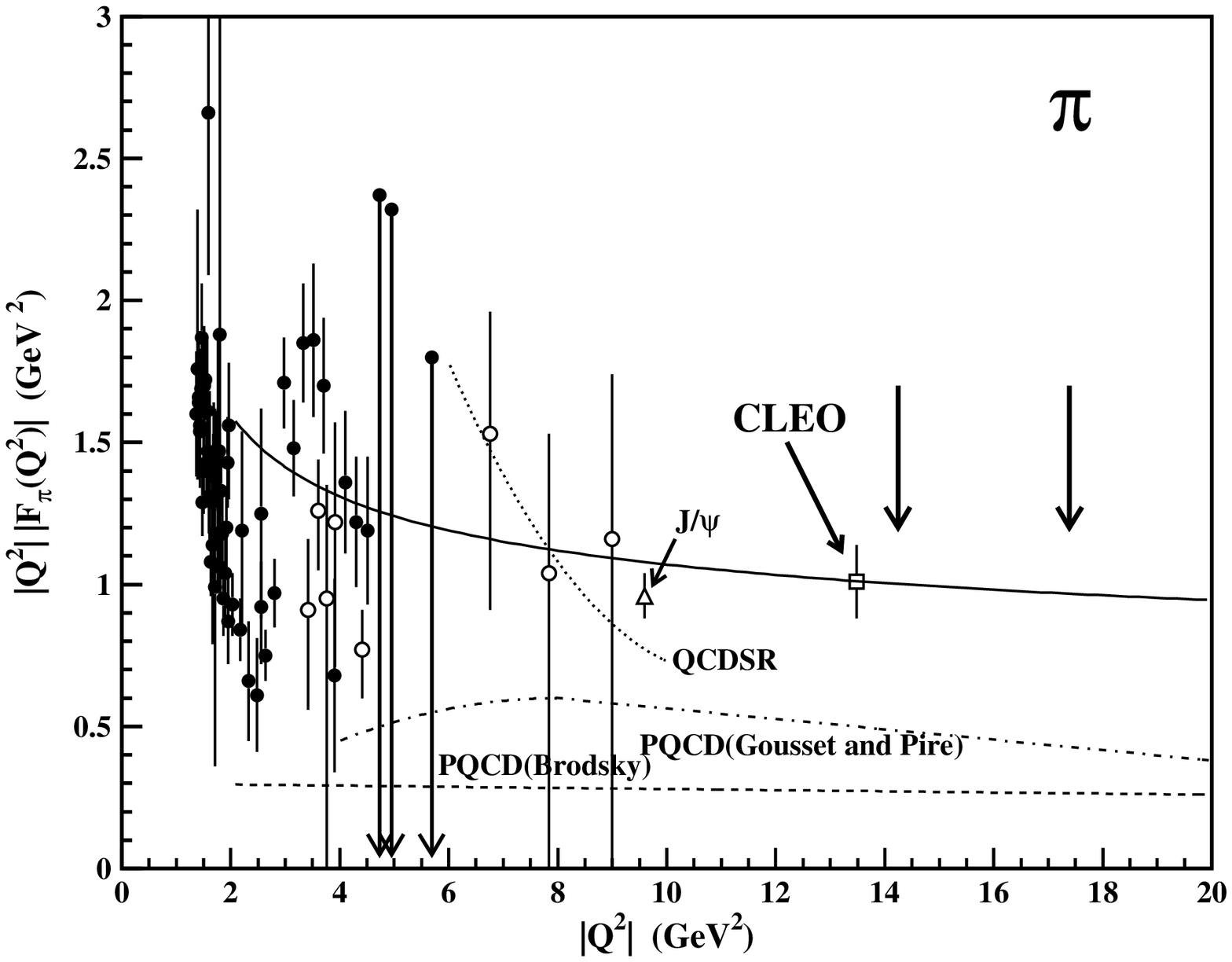}
\includegraphics[width=2.9in]{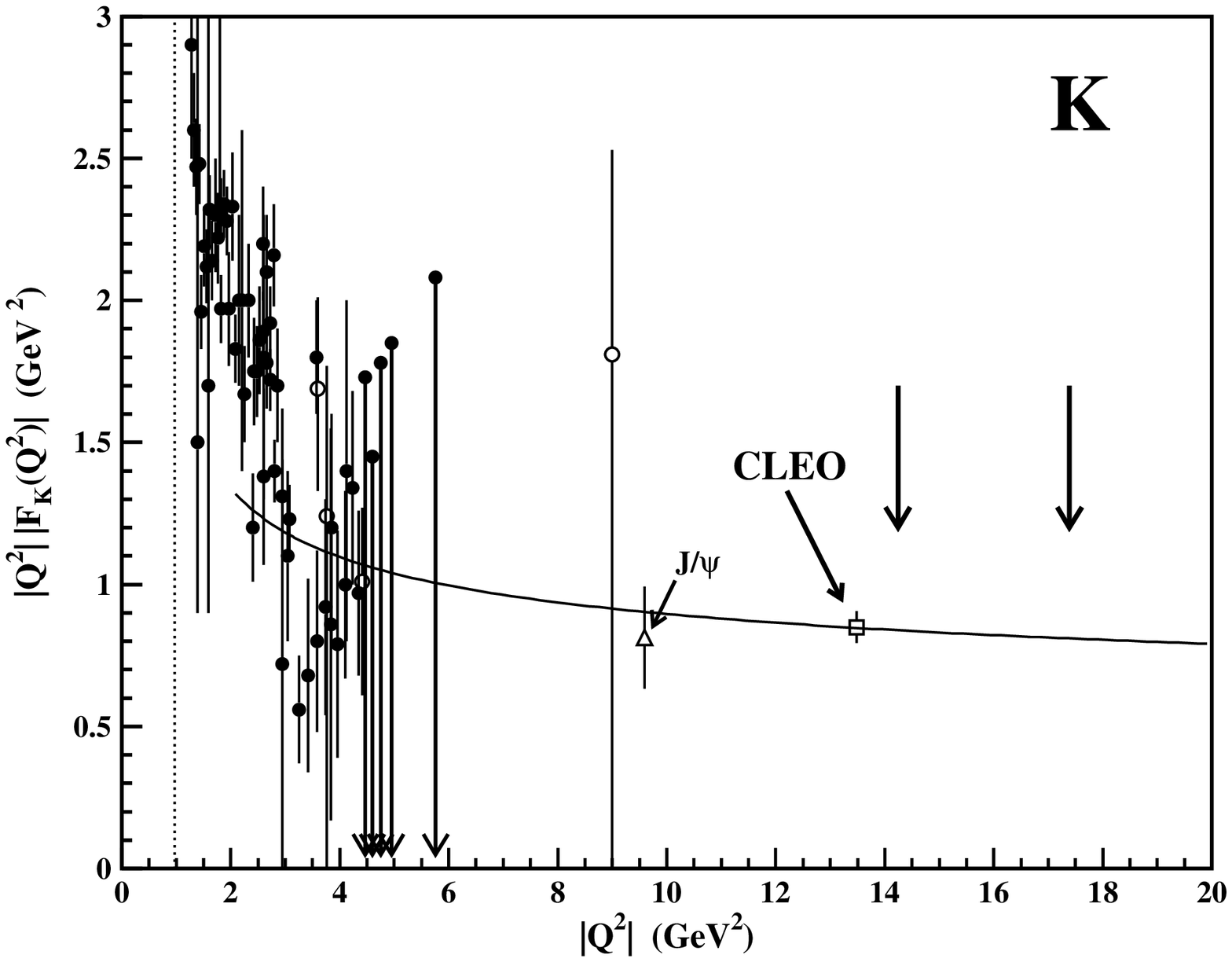}
\end{center}
\caption{Pion and kaon timelike form factors including CLEO published results.  Arrows mark where new CLEO results are expected.  The theoretical predictions available for pions are also shown.}
\end{figure}

We note that the experimental ratios for hadronic to leptonic decays of $J/\psi$ and $\psi'$ are nearly the same, $R(p\bar{p}/e^+e^-)\approx3.7\times10^{-2}$, $R(K^+K^-/e^+e^-)\approx6\times10^{-4}$, $R(\pi^+\pi^-/e^+e^-)\approx3\times10^{-4}$.  If we \textbf{assume} that these ratios remain the same for $\psi(3770)$ and $\psi(4160)$ we can use the measured $\mathcal{B}(\psi(3770,4160)\to e^+e^-)$ to estimate the branching fractions for the decay of these resonances to obtain $\mathcal{B}(\psi(3770,4160)\to p\bar{p})\approx4\times10^{-7}$,  $\mathcal{B}(\psi(3770,4160)\to \pi^+\pi^-)\approx3\times10^{-9}$, and  $\mathcal{B}(\psi(3770,4160)\to K^+K^-)\approx6\times10^{-9}$.  These lead to estimated resonance cross sections of $\sim4$~fb~($p\bar{p}$), $\sim0.3$~fb~($\pi^+\pi^-$), $\sim0.6$~fb~($K^+K^-$).  If the measured cross sections turn out to be \textbf{substantially larger} than these, they can be attributed to form factor contributions.  In other words, we can obtain $G_M(p\bar{p})$, $F_\pi$, $F_K$ at $Q^2=14.2$ and 17.3~GeV$^2$ with much better precision than that obtained at $Q=13.45$~GeV$^2$.  Counts in the hundreds are expected.  The arrows in Fig.~3 indicate where these measurements will sit on the $Q^2F(\pi\pi,KK)$ plots.  Stay tuned for the results.

In Fig.~3, we also show the theoretical predictions for $Q^2F_\pi$.  Needless to say, none of the predictions come even close to the precision experimental results.  Since there is no hope that lattice calculations can shed light on timelike form factors (they work in Euclidean time), it is a big challenge and opportunity for non--lattice theorists.

\end{document}